\documentclass[useAMS,usenatbib,usegraphicx]{mn2e}
\usepackage{url}

\def\kperp{{k_\perp}}
\def\kpara{{k_\parallel}}

\begin{document}

\title[New limits on the H{\small I} power spectrum from GMRT]
{A simulation calibrated limit on the H{\LARGE I} power spectrum from the GMRT Epoch of Reionization experiment}

\author[G. Paciga et al.]
 {Gregory~Paciga$^1$\thanks{Email:paciga@astro.utoronto.ca},
   Joshua~G.~Albert$^1$,
   Kevin~Bandura$^{2,3}$,
   Tzu-Ching~Chang$^{1,4}$,
   \newauthor
   Yashwant~Gupta$^5$,
   Christopher~Hirata$^6$,
   Julia~Odegova$^1$,
   Ue-Li~Pen$^1$\thanks{Email:pen@cita.utoronto.ca},
   \newauthor
   Jeffrey~B.~Peterson$^2$,
   Jayanta~Roy$^5$,
   J.~Richard~Shaw$^1$,
   Kris~Sigurdson$^7$,
   \newauthor
   Tabitha~Voytek$^2$\\
   $^1$ CITA, University of Toronto, 60 St George Street, Toronto, ON M5S 3H8, Canada \\
   $^2$ Department of Physics, Carnegie Mellon University, 5000 Forbes Ave, Pittsburgh, PA 15213, USA \\
   $^3$ Department of Physics, McGill University, 3600 rue University, Montr\'{e}al, QC H3A 2T8, Canada \\
   $^4$ IAA, Academia Sinica, PO Box 23-141, Taipei 10617, Taiwan \\
   $^5$ National Center for Radio Astrophysics, Tata Institute for Fundamental Research, Pune 411 007, India \\
   $^6$ Department of Astrophysics, Caltech M/C 350-17, Pasadena CA 91125, USA \\
   $^7$ Department of Physics and Astronomy, University of British Columbia, Vancouver, BC V6T 1Z1, Canada \\
}

\maketitle

\begin{abstract}
The Giant Metrewave Radio Telescope Epoch of Reionization experiment is an ongoing effort to measure
the power spectrum from neutral hydrogen at high redshift.
We have previously reported an upper limit of (70\,mK)$^2$
at wavenumbers of $k \approx 0.65\,h\,\mathrm{Mpc}^{-1}$
using a basic piecewise-linear foreground subtraction.
In this paper we explore the use of a singular value decomposition
to remove foregrounds with fewer assumptions about the foreground structure.
Using this method we also quantify, for the first time,
the signal loss due to the foreground filter
and present new power spectra adjusted for this loss,
providing a revised measurement of
a $2\sigma$ upper limit at (248\,mK)$^2$
for $k = 0.50\,h\,\mathrm{Mpc}^{-1}$.
While this revised limit is larger than previously reported,
we believe it to be more robust and still represents
the best current constraint on reionization at $z \approx 8.6$.
\end{abstract}


\section{Introduction}
\label{sec:Introduction}

The Epoch of Reionization (EoR) began as the first stars ionized the neutral hydrogen
around them, and ended when that ionization extended across most of the Hubble sphere.
\citet{Furlanetto06} provides a thorough review of the subject.
Based on the electron column density to the cosmic microwave background (CMB)
and under the assumption that reionization was instantaneous,
\textit{Wilkinson Microwave Anisotropy Probe} (\textit{WMAP}) data suggests it would have occurred at $z=10.4$ \citep{Komatsu11}. 
Theoretical work, however, has generally suggested that reionization
was a patchy and extended process
\citep{Furlanetto04, McQuinn07, Zahn07, Friedrich11, Su11, Griffen12}.
Observations of absorption lines in quasar spectra 
can be used to limit the fraction of neutral hydrogen at high
redshift \citep{GunnPeterson65}
and it is generally accepted that
reionization was complete by a redshift of $z\approx 6$
\citep{Becker01, Djorgovski01},
though the actual H{\small I} fraction may still have been quite high
\citep{McGreer11, Schroeder12}.
Using the global 21\,cm signal as a function of redshift, \citet{Bowman10}
have put a lower limit on the duration of the EoR of $\Delta z > 0.06$,
while \citet{Zahn12} have used 
measurements of the kinetic Sunyaev--Zel'dovich effect
with the South Pole Telescope 
to suggest an upper limit on the transition
from a neutral fraction of 0.99 to 0.20 of $\Delta z < 4.4$.

The redshifted 21\,cm H{\small I} spectral line can be used
to trace the patchy distribution of neutral hydrogen in the Universe before
the first luminous sources formed until the end of the EoR \citep{Furlanetto06}.
The distribution during the transition can be used to
constrain cosmological parameters
\citep{Furlanetto09, Mao08, Cooray08, McQuinn06, Pandolfi11} 
and deduce the nature of the first ionizing sources themselves
\citep{Iliev12, Kovetz12, Datta12, Majumdar12},
including possible exotic reionization scenarios
\citep[e.g.,][]{Furlanetto06BHA, Haiman11}.
Unfortunately the 21\,cm signal is many orders of magnitude less 
than foregrounds from Galactic and extragalactic sources 
at the relevant frequencies \citep{Oh03, dOC08}.
These foregrounds are currently one of the largest obstacles
to detecting the 21\,cm signal and
several schemes have been developed to address the problem 
\citep[e.g.,][]{Petrovic11, Liu11, Chapman12, Dillon12, Parsons12}.

Several groups are making progress towards measuring the 21\,cm power spectrum.
\citet{Bowman08} estimated an upper limit to
the contribution of H{\small I} to the redshifted 21\,cm brightness temperature
of 450\,mK.
The Precision Array for Probing the Epoch of Reionization
(PAPER) has reported a limit
of approximately 5\,K with a 310\,mK noise level \citep{Parsons10}.
The Murchison Widefield Array \citep[MWA;][]{Lonsdale09}
is expected to be able to detect both the amplitude
and slope of the power spectrum with a
signal-to-noise ratio $>10$ \citep{Beardsley12}.
Both PAPER and MWA have emphasized the importance
of antenna layout in maximizing sensitivity to the EoR
signal \citep{Parsons11, Beardsley12}.
The Low-Frequency Array (LOFAR) names the EoR as
one of its Key Science Projects \citep{Harker10, Brentjens11}
and is currently being commissioned in the Netherlands.
\citet{Zaroubi12} estimate that LOFAR
will have the potential to overcome the low signal-to-noise
to directly image the neutral hydrogen.
Future generations of telescopes,
in particular the Square Kilometre Array,
should be capable of direct imaging \citep{Carilli04}
but will not be in full operation for
another decade \citep{Rawlings11}.

The GMRT-EoR experiment has been an ongoing effort
using the Giant Metrewave Radio Telescope (GMRT) in India \citep{Swarup91, GMRT},
which in contrast to other experiments features
large steerable antennas with a collecting area comparable to LOFAR,
and a relatively small field of view.
In \citet{Paciga11} we reported an upper limit on the neutral hydrogen
power spectrum of (70\,mK)$^2$ at $2\sigma$ using a simple
piecewise-linear foreground filter.
However, this limit did not account for any 21\,cm signal lost
in the foreground filter itself. The purpose of the
current work is to quantify the potential signal loss,
and to present these results with a new singular value decomposition (SVD) foreground filter.

This paper is organized as follows.
In section~\ref{sec:data} we briefly describe the data and the 
preliminary data analysis.
In section~\ref{sec:foreground-removal} we discuss the SVD foreground
filter, followed by quantifying the signal loss it causes.
Finally,
we make an estimate of the full 3D power spectrum in section~\ref{sec:3dpower}
and conclude in section~\ref{sec:conclusion}.
When necessary we will use the \textit{WMAP}7 maximum likelihood 
parameters
$\Omega_\mathrm{M}=0.271$,
$\Omega_\Lambda = 0.729$ and $H_0 = 70.3\,\mathrm{km\,s^{-1}\,Mpc^{-1}}$
\citep{Komatsu11}.
All distances are in comoving units.

\begin{figure}
  \includegraphics[width=\columnwidth]{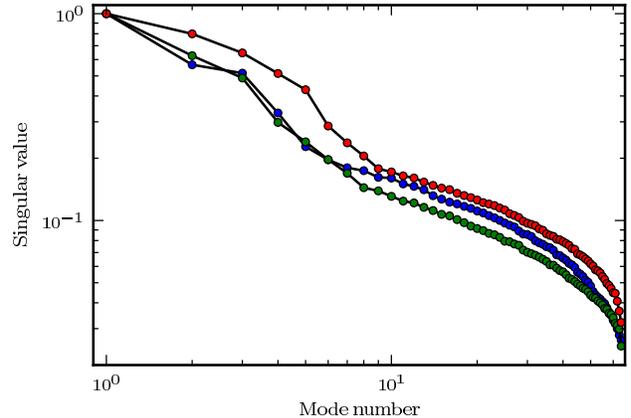}
  \caption{An example singular value spectrum for the three shortest
    baselines, each approximately 50--52 wavelengths at the zenith,
    with the largest singular value normalized to 1 for each.}
  \label{fig:eigenspectra}
\end{figure}

\begin{figure*}
  \includegraphics[width=\textwidth]{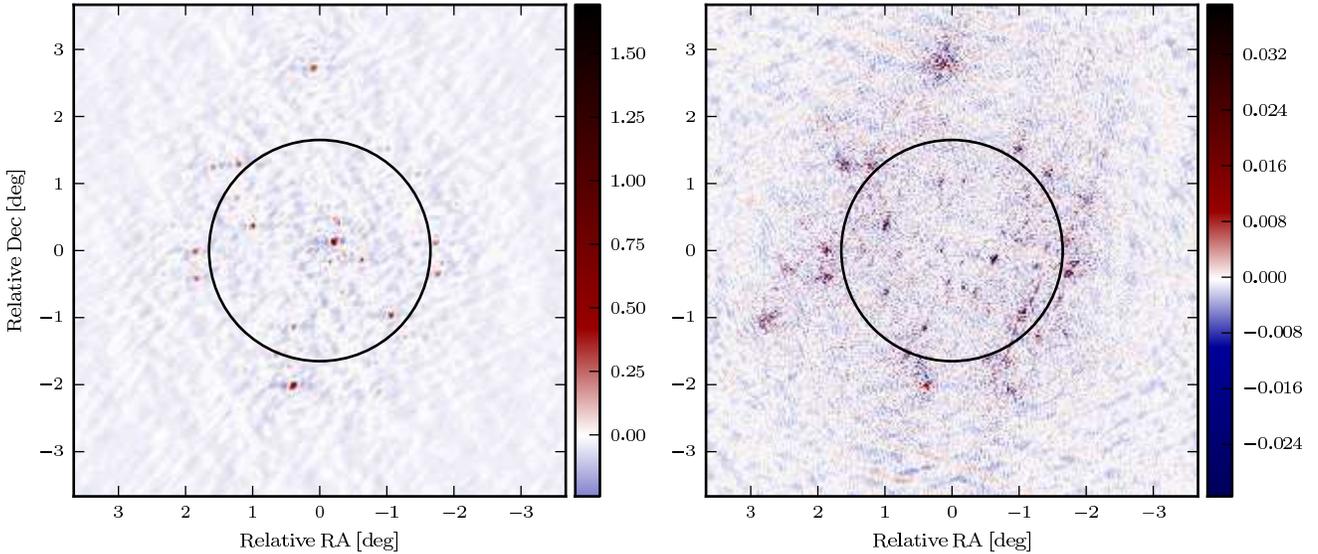}
  \caption{Sky image before and after an SVD foreground subtraction,
    for the night of 2007 December 10, using baselines up to 4\,km.
    The colour scales are in units of Janskys and the black circle
    shows the full width half-maximum (FWHM) of the primary beam.
    The rms before any foreground
    removal (left) is 50\,mJy. After removing eight SVD modes (right)
    the peak goes from 1.6\,Jy to 39\,mJy with an rms of 2\,mJy.
    Residual point sources can still be seen around the edge of the beam
    while those within about one degree of the centre are effectively removed.
  }
  \label{fig:sky}
\end{figure*}

\begin{figure*}
  \includegraphics[width=\textwidth]{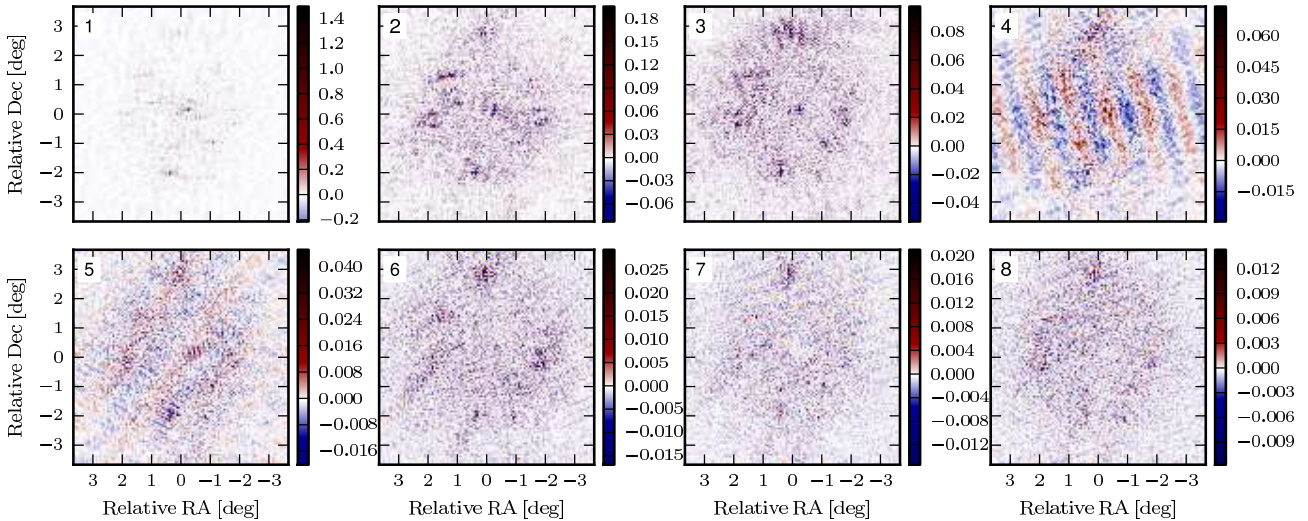}
  \caption{Sky images of the top eight SVD modes identified in the data from 2007 December 10,
  with all other modes set to zero.
  These modes are the ones subtracted between the two sky images in Fig.~\ref{fig:sky}.
  The colour scales are in units of Janskys.}
  \label{fig:eigenmodes}
\end{figure*}

\section{Observations and Data}
\label{sec:data}

The data analysed in this paper were taken over five
nights in 2007 December and total about 40\,h.
The observations were centred on PSR~B0823+26,
a pulsar about 30$^\circ$ off the Galactic plane,
which is used to calibrate both the phases and the
ionosphere by dividing the pulsar period into several gates
and separating the on- and off-pulse gates.
The primary beam has a full width half-maximum (FWHM) of 3.1$^\circ$
and a maximum angular resolution of about 20~arcsec.
The bandwidth covers a frequency range from 139.3 to 156.0\,MHz
in 64 frequency bins of 0.25\,MHz each
with a time resolution of 64\,s.
This corresponds to a redshift range of $z=8.1$--$9.2$.
For more details on the observations, including radio frequency interference (RFI) removal strategies,
see \citet{Paciga11}.
The remainder of this section outlines the differences
in the data analysis compared to this earlier work.

In addition to automated flagging of
visibilities and the SVD RFI removal pipeline
for broad-band interference, 
we have also added manual flagging
of faulty antennas, timestamps, and frequency ranges
that are exceptionally noisy
or that visually appear to have RFI left
after the automated procedures.
Approximately 15 per cent of the visibilities are flagged in this way,
while the dynamic range,
defined as the ratio of the peak flux of the calibration source
to the rms outside the primary beam,
is improved by as much as a factor of 4.

To improve the comparability of each of the five nights of observing, 
we limit each night to the same LST range,
and have regridded the visibilities in time
such that each night shares the exact
same timestamps, equally spaced in one minute
intervals.
Additionally,
it is known that the flux of a pulsar
can change significantly with time,
which creates another source of variability from night to night.
Since the visibilities are gated on the period of the pulsar,
it can easily be subtracted from the data.
These two changes reduce the rms noise
in the difference between pairs of days
by a factor of 2 on average.

Finally, the power spectrum is calculated from the
cross-correlation of pairs of nights
in annuli of $(u,v)$ space.
This gives the 2D power perpendicular
to the line of sight as a function of baseline length $|\mathbf{u}|$
or equivalently multipole moment $\ell = 2\pi|\mathbf{u}|$.
At the smallest $|\mathbf{u}|$, a bin width equal to that
of the primary beam (20 wavelengths) was used. At larger
$|\mathbf{u}|$, each bin width is increased by 60 per cent,
 to compensate for the decreasing density of visibilities.

It was found that outliers tended to skew
the mean power in each annulus.
Since the median 
is much more robust to such outliers,
we calculate the power in each annulus (that is, at each angular scale $\ell$) as
the median value over all frequencies.
The error in each annulus is estimated
as the median of the absolute deviations from the median power,
weighted by the noise.
The final power spectrum is
the bootstrapped average
of the power spectra
over all 10
possible cross-correlation pairs of the 5 nights.

The power can also be expressed in terms of
the wavenumber perpendicular to the line of sight,
$\kperp \approx (\ell / 6608 )\,h\,\mathrm{Mpc}^{-1}$,
which becomes useful when discussing the 3D power.
In the 2D case we will continue to use $\ell$.
Since we do not yet include line-of-sight information,
this is the power as a function of $\ell$ with fixed $\kpara=0$,
which we denote $P(\ell | \kpara=0)$.

\section{Foreground Removal}
\label{sec:foreground-removal}

\subsection{Singular Value Decomposition}
\label{sec:svd}

Foreground removal techniques typically rely on the fact that
the foreground signal is expected to be much smoother
in frequency 
(that is, has fewer degrees of freedom)
than the reionization signal,
which decorrelates on the order of one to a few megahertz \citep{Bharadwaj05}.
Observations of foregrounds around 150\,MHz with GMRT have
shown that the fluctuations in frequency are large
enough to make polynomial fits insufficient to model them \citep{Ali08, Ghosh12}.
In this work, we instead use an SVD,
which still isolates smooth foreground modes but does not
make a priori assumptions that the foregrounds 
can be approximated with a particular function.
A similar technique has been used by \citet{Chang10} and \citet{Masui12}
to clean foregrounds for H{\small I} intensity mapping at $z\approx0.8$,
where the relative dominance of foregrounds over the 21\,cm signal
is comparable to $z\approx8.6$.
For reionization, \citet{Liu12} have developed a framework
for using SVD modes of a frequency-frequency correlation matrix
to clean foregrounds at MWA.

We perform an SVD for each baseline individually
on the visibilities arranged in a matrix by time and frequency.
The number of modes is limited by the 64 frequency channels.
Fig.~\ref{fig:eigenspectra} shows the singular values for the
shortest baselines. The spectra of values on a given baseline
is generally consistent from day to day, but occasionally there are
large jumps in both amplitude and rate of decline with mode number,
which are likely due to either RFI or calibration errors.
In these cases, the noise on the baseline also becomes
much larger, such that in the final calculation of the power spectrum
their contribution is significantly down-weighted.

A sky image using 8\,h of data
from a single night is shown in Fig.~\ref{fig:sky},
compared with the same data after the first eight SVD modes,
shown in Fig.~\ref{fig:eigenmodes}, are removed.
The overall flux is reduced substantially after only a few
modes are removed. While the sources in the centre
of the field are removed quite well, the dominant
residuals are the point sources near the edge of the beam.
This is generically true of any foreground subtraction
used on this data set, as was also seen in \citet{Paciga11}.
This is most likely due to beam edge effects,
the worst residuals being close to the first null
where the frequency dependence of the beam pattern
is most significant.
Though there are sophisticated schemes that may be
able to model point sources while minimizing the impact on the 21\,cm signal
\citep[e.g.,][]{Datta10, Bernardi11, Trott12},
at the angular scales we are interested in for this work
($\ell \la 2000$) the point sources are confusion limited
and contribute in the same way as the diffuse background.

\begin{figure}
  \includegraphics[width=\columnwidth]{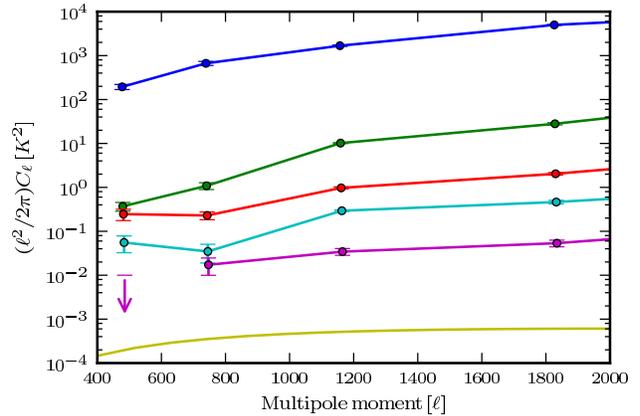}
  \caption{Power spectra before and after SVD mode removal.
  The blue line shows before any modes are removed.
  The green, red, cyan, and purple lines are for
  4, 8, 16, and 32 modes removed, respectively.
  The error bars are
  from a bootstrap analysis of all cross-correlated
  pairs from the five nights of data.
  The solid yellow line represents the theoretical signal
  from \citet{Jelic08}.}
  \label{fig:psvd}
\end{figure}

Each night goes through the SVD foreground removal
separately, and then the cross-correlations
are used to arrive at a power spectrum
using the method described in section~\ref{sec:data}. 
The spectra for several numbers of SVD modes
removed are shown in Fig.~\ref{fig:psvd}.

\subsection{Quantifying Signal Loss}
\label{sec:quantifying-loss}

A general problem with any foreground removal strategy is that it is
impossible to completely separate the foregrounds from the signal,
such that the foreground removal will likely remove some signal as well.
Early work by \citet{Nityananda10} used a simple model
of an SVD applied to a single visibility matrix
to show that the signal loss could be calculated analytically.
Our method of using an SVD for each baseline independently is more complex,
and we wish to estimate the signal loss directly from the data itself.
To quantify the signal loss, we aim to find the transfer function
between the observed power $P_\mathrm{SVD}(\ell)$
and the real 21\,cm power $P_{21\mathrm{cm}}(\ell)$.
Since the real power is unknown, we use a simulated signal as a proxy.
This is added to the data before the foreground subtraction and
the resulting power spectrum after subtraction is compared to the input signal.

\begin{figure}
  \includegraphics[width=\columnwidth]{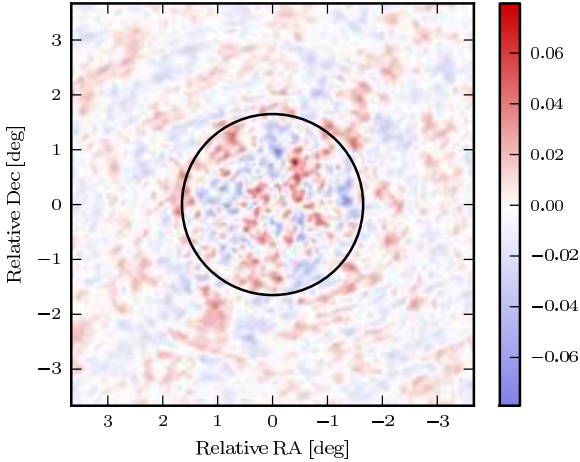}
  \caption{An image of the simulated signal as seen with GMRT
    in the absence of foregrounds or noise, using the
    same baselines and field of view as Fig.s~\ref{fig:sky}
    and~\ref{fig:eigenmodes}. The colour scale is in units of mJy.
  The solid circle represents the FWHM of the primary beam.}
  \label{fig:simsky}
\end{figure}

The simulated signal we use is a Gaussian random field
with a matter overdensity power spectrum from {\small CAMB}\footnote{Code for Anisotropies in the Microwave Background; \url{http://camb.info}.}
scaled to $z=8.6$ using the linear-regime growth from $z=1.5$,
and with the amplitude calibrated to be similar to
the expected 21\,cm signal from EoR
assuming that the spin temperature is much greater than the CMB temperature.
Fig.~\ref{fig:simsky} is an image of the simulated signal
as it would be seen by GMRT
in the absence of any foregrounds or noise.
The effect of the beam profile on the power spectrum
is less than 3 per cent  for scales in the range $40<\ell<2000$,
and so has a relatively small effect on the result.

Given a data set $x$ which is the sum of the observed data
and a simulated signal, the transfer function
$T(x \rightarrow F(x); \ell)$
 measures
how much of the signal survives in the foreground filtered
data set $F(x)$ as a function of $\ell$.
While $F(x)$ can stand in for any filter method applied to the visibilities $x$,
for the SVD we must also specify that the modes removed are those
calculated from the visibilities $x$ themselves.
While the transfer function measures the signal loss
for a single set of data, the power is measured from the
cross-correlations of those data sets,
so the relationship can be written as
\begin{equation}
\label{eqn:Tgeneral}
P_\mathrm{SVD}(\ell) = T(x \rightarrow F(x); \ell)^2 P_{21\mathrm{cm}}(\ell) .
\end{equation}
\noindent 
Unless it is necessary
to be explicit about the mapping $T$ is measuring,
we will shorten this notation to simply $T(\ell)$.

There are numerous ways one can estimate this function.
The most direct way is to cross-correlate $F(x)$ with
the injected signal, and normalize by the auto-power of that same signal.
This is written as
\begin{equation}
\label{eqn:T0}
T_0(\ell)
=  \frac{F(\textrm{data} + \textrm{signal}) \times \textrm{signal}}
                  {\textrm{signal} \times \textrm{signal}} .
\end{equation}
\noindent 
In the ideal case where $F(x)$ removes foregrounds perfectly
this will equal exactly 1.0.
While conceptually simple, 
and used successfully by \citet{Masui12}
for data at $z\approx0.8$,
we find
this estimator
of the transfer function to be exceptionally noisy
for realistic cases where $F(x)$ leaves
residual foregrounds. In the case of the SVD,
we would expect the function to become less noisy as more modes are removed
and the residual foregrounds decrease,
but we are still significantly limited in being able to measure the power.

An alternative is to subtract the original visibilities,
under the same foreground filter,
from the combined real and simulated visibilities
before cross-correlating with the simulated signal.
To distinguish it from the previous one, we denote this 
version of the transfer function $T_1$,
which takes the form
\begin{equation}
\label{eqn:T1}
T_1(\ell) = \frac{[F(\textrm{data} + \textrm{signal})  - F(\textrm{data})] \times \textrm{signal}}
           {\textrm{signal} \times \textrm{signal}} .
\end{equation}
\noindent In addition to being much less
noisy when residual foregrounds are present,
this has the benefit that by subtracting
the original data we remove the possibility of the real 21\,cm signal in
the data correlating with the simulated signal and biasing the result.
If $F(x)$ left the signal untouched,
this would in principle be equal to 1.0.
However, deviations are possible even when $F(x) = x$.
This is due to the fact that the cross-correlations with real data in the numerator
introduces RFI masks, noise, and day-to-day variations
which are not present in the pure signal in the denominator. 
Thus, the transfer function will also correct for these effects,
which enter at a level of a few per cent.

\begin{figure}
  \includegraphics[width=\columnwidth]{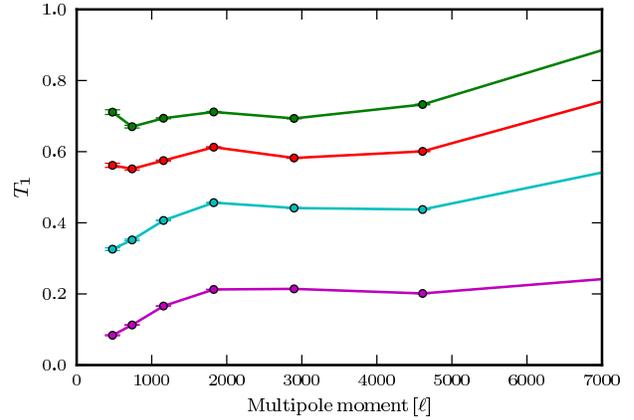}
  \caption{Transfer function $T_1$ 
    with 
    4 (green), 8 (red), 16 (cyan), and 32 (purple) SVD modes removed,
    showing the fraction of the 21\,cm signal that we estimate survives
    the SVD foreground removal. 
    With only
    four modes removed, most of the 21\,cm signal is expected to survive.
    However, when 32 modes are removed, about 20 per cent or less survives,
    depending on the angular scale.}
  \label{fig:tfunc}
\end{figure}

We carry out this process of estimating $T(\ell)$,
averaging over 100 realizations of the simulated signal,
after which both the mean and the standard deviation are well determined,
and the error in the mean is small enough that it will not contribute
significantly to the corrected power spectra later.
Fig.~\ref{fig:tfunc} shows $T_1$ for a selection
of SVD filters.
While the transfer function in principle can depend non-linearly
on the amplitude of the input signal, we find that the
result does not change significantly
within a factor of 10 of realistic signal
temperatures.
In the regimes where the transfer function does begin
to depend on the input temperature, the two are anti-correlated;
larger signals are more readily misidentified
as foregrounds by the SVD,
leading to a small value of $T(\ell)$.

\begin{figure}
  \includegraphics[width=\columnwidth]{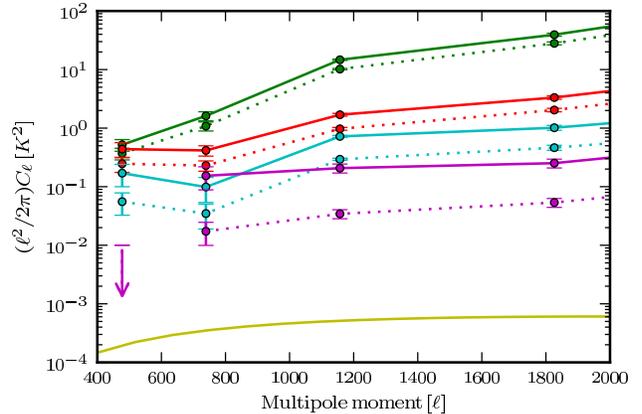}
  \caption{The $T_1$-corrected power spectra after SVD foreground removal.
    The colours represent 4 (green), 8 (red), 16 (cyan), and 32 (purple)
    SVD modes removed.
    The dotted lines show the uncorrected power spectra as in Fig.~\ref{fig:psvd},
    while the solid lines show the power spectra after correcting
    for the transfer function.
    For 32 modes removed, at low $\ell$,
    the corrected power is larger than that for only 16 modes removed.
    As in Fig.~\ref{fig:psvd}, the solid yellow line represents the theoretical
  signal from \citet{Jelic08}.}
  \label{fig:psvd-corrected}
\end{figure}

The transfer function can be used to determine the best
number of modes to remove, since as more modes are removed
more of the 21\,cm signal will be reduced to a point where
the additional correction to the signal outweighs the gain
from reducing the foregrounds. Fig.~\ref{fig:psvd-corrected}
shows that correcting for the transfer function after 32 modes
are removed gives a weaker limit on the power than only
removing 16 modes.

\section{3D Power Spectrum}
\label{sec:3dpower}

\subsection{Line-of-sight power}
\label{sec:hermite}

The power calculated from annuli in visibility space only measures
the 2D power perpendicular to the line of sight
(that is, as a function of the multipole moment $\ell$ or wavenumber $\kperp$).
To find the full 3D power, we must also look at the line-of-sight,
or frequency, direction
and measure power as a function of $\kpara$.
While certain forms of foreground filters will have a window function
that naturally selects a $\kpara$, the SVD filter does not have
a well defined behaviour along the line of sight.
The gives us the flexibility of selecting the window function.

\begin{figure}
\includegraphics[width=\columnwidth]{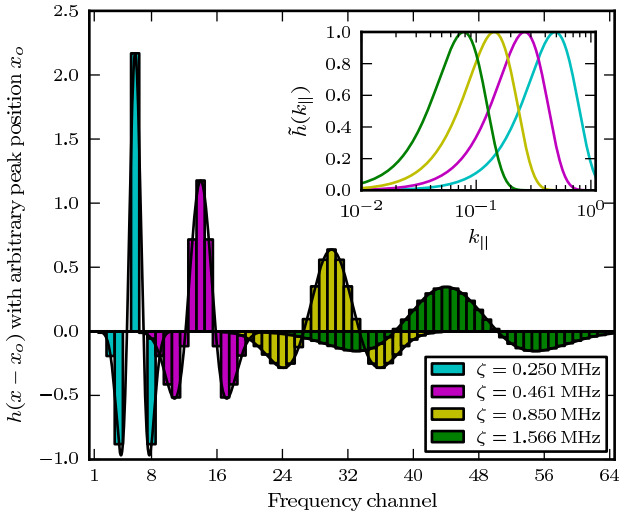}
\caption{Hermite window in frequency space. Four examples are shown with
different values of $\zeta$ increasing from left to right, with arbitrary
horizontal offset. Bars indicate the value of the window in each frequency bin.
In practice, $\zeta$ should not be smaller than the frequency resolution,
limiting the $\kpara$ available.
The inset plot shows the Fourier transform.}
\label{fig:hermite}
\end{figure}

Hermite functions, having the benefit of zero mean
and a simple Fourier transform, are well suited to
select a range of $\kpara$.
In frequency space, we define a window
\begin{equation}
\label{eqn:hermite-nu}
h(\nu) = \frac{1}{\sqrt{8\pi} \zeta} \left( 1 - \frac{\nu^2}{\zeta^2} \right) \exp \left( 1 - \frac{\nu^2}{2\zeta^2} \right)
\end{equation}
\noindent where $\zeta$ is a parameter analogous to the
standard deviation of a Gaussian distribution,
which in this case specifies the location of the zeros.
This is shown in Fig.~\ref{fig:hermite} for several $\zeta$ compared to the
frequency bin size.
This window has the Fourier transform
\begin{equation}
\label{eqn:hermite-k}
\tilde{h}(\kpara) = \frac{(\kpara r \zeta)^2}{2} \exp \left[ 1 - \frac{(\kpara r \zeta)^2}{2} \right ] .
\end{equation}
\noindent We have used the conversion factor
$r\approx11.6\,h^{-1}\mathrm{Mpc}/\mathrm{MHz}$
such that $\kpara$ is in units of $h\,\mathrm{Mpc}^{-1}$.
The normalization has been chosen such that
the maximum of $\tilde{h}(\kpara)$ is 1, thus preserving power.
This peak in Fourier space, shown in the inset of Fig.~\ref{fig:hermite},
occurs when $\kpara = \sqrt{2}/(r\zeta)$ and
determines the $\kpara$ at which most power survives the Hermite window.
Larger $\zeta$ sample smaller $\kpara$,
with the range of possible values limited by the frequency
resolution and bandwidth.

\subsection{Three approaches to the transfer function}

\begin{figure}
  \includegraphics[width=\columnwidth]{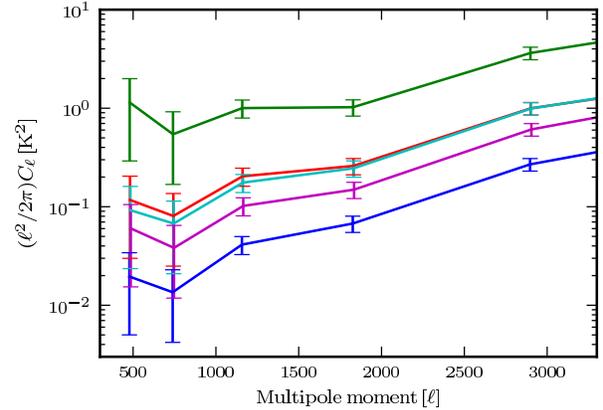} 
  \caption{An example power spectrum at $\zeta = 0.25$\,MHz
    ($\kpara = 0.49\,h\,\mathrm{Mpc}^{-1}$)
    with 16 SVD modes removed (blue line), 
    with the effect of each type of
    transfer function correction for $T_1$ shown.
    The green line corrects for both the Hermite window
    and the SVD subtraction, while the red line reintroduces the
    Hermite window,
    which agrees quite well with the semi-Hermite correction (cyan).
    The purple line uses the
    SVD-only transfer function.
    Error bars are from include contributions from the transfer function
    and the bootstrap error from the raw power spectra.
    In this example, the three approaches agree quite well,
    though they can diverge by an order of magnitude
    for other selections of mode subtraction and $\kpara$.
}
  \label{fig:transfer-example}
\end{figure}

By applying a Hermite window to the data,
we can calculate the 2D power spectrum at a fixed $\kpara$.
There is some complication, however, in how we can use a 
transfer function to correct for possible signal loss.
The Hermite filter by design reduces power on most scales
while leaving power only at a specific $\kpara$,
and we want the transfer function to only adjust for signal
lost at the same $\kpara$. 
Ideally, one would apply the Hermite filter first to isolate
the input power at the scales of interest and run the
foreground filters on that data.
If $H(x)$ represents the data set with a Hermite window applied,
this would measure $T(H(x) \rightarrow F(H(x)); \ell)$.
Unfortunately, the SVD actually
depends strongly on information in the $\kpara$ direction,
which means that $F(H(x))$ may have a much different effect
on the power at the chosen length scale than $F(x)$.
That is to say, the Hermite and SVD operations do not commute.

There are several possible approaches to get around this, which are as follows:
\begin{enumerate}

\item Assume that the transfer function is not
strongly dependent on $\kpara$, and use 
$T(x \rightarrow F(x); \ell)$ from the $\kpara=0$ case
independent of the $\kpara$ selected by the Hermite window.
We call this the `SVD only' approach.
The $\kpara$ behaviour only enters in through 
calculation of the power spectra after the Hermite window.
Any important behaviour of the SVD in the $\kpara$ direction
will not be captured.

\item 
We can calculate a transfer function for the signal
loss due to the total effect of both the Hermite window and the SVD,
$T(x \rightarrow H(F(x)); \ell)$,
and correct for both. 
We can then use an analytical form for the transfer
function of the Hermite window alone
to reintroduce the scale window and keep only the power
at our selected $\kpara$.
We call this the `full Hermite' approach.

To find its analytical form, we
start with the fact that the transfer function associated with the Hermite
window measures the ratio of the windowed power
to the full power,
\begin{equation}
T_{H}^2(\kperp) =
\frac{ \int P(\kperp, \kpara) | \tilde{h}(\kpara) |^2 \mathrm{d}\kpara }
     { \int P(\kperp, \kpara) \mathrm{d} \kpara } .
\end{equation}
If we assume the power spectrum has the form
\begin{equation}
P(\kpara) \propto \frac{1}{\kpara^2 + \kperp^2}
\end{equation}
\noindent both the numerator and denominator of this
can be evaluated analytically. The result is
\begin{eqnarray}
\label{eqn:analytical-hermite}
T_{H}^2(\kperp) &=&\frac{e^2 r\zeta\kperp}{8\sqrt{\pi}}
   \left( 1 - 2r^2\zeta^2\kperp^2 \phantom{e^{r^2 \zeta^2 \kperp^2}} \right. \nonumber \\
&& \left. + 2\sqrt{\pi}r^3\zeta^3 \kperp^3 e^{r^2 \zeta^2 \kperp^2} \mathrm{erfc}[r\zeta\kperp] \right)
\end{eqnarray}
\noindent where $\mathrm{erfc}[x] = 1 - \mathrm{erf}[x]$
is the complimentary error function.
Requiring the most steps, this method has
more avenues to introduce errors or biases.

\item Apply the Hermite window first to the simulated signal.
When added to the full data and passed through the SVD foreground
removal, the larger amplitude of the foregrounds present in the data
ensures that the SVD still has data at all $\kpara$ to operate on.
However, since there is only a simulated signal at a specific
$\kpara$, the cross-correlation with the simulated signal
when calculating the transfer function 
$T(\mathrm{data} + H(\mathrm{signal}) \rightarrow F(\mathrm{data} + H(\mathrm{signal})); \ell)$
only measures the effect
of the SVD on that $\kpara$.
We call this the `semi-Hermite' approach.
This assumes that the SVD as applied to the 
$\kpara$ limited simulated signal is a suitable
proxy for how the SVD affects the real signal,
given that both the real signal and the $\kpara$ limited simulated
signal are of significantly lower amplitude than the foregrounds.

\end{enumerate}

\begin{figure*}
\includegraphics[width=\textwidth]{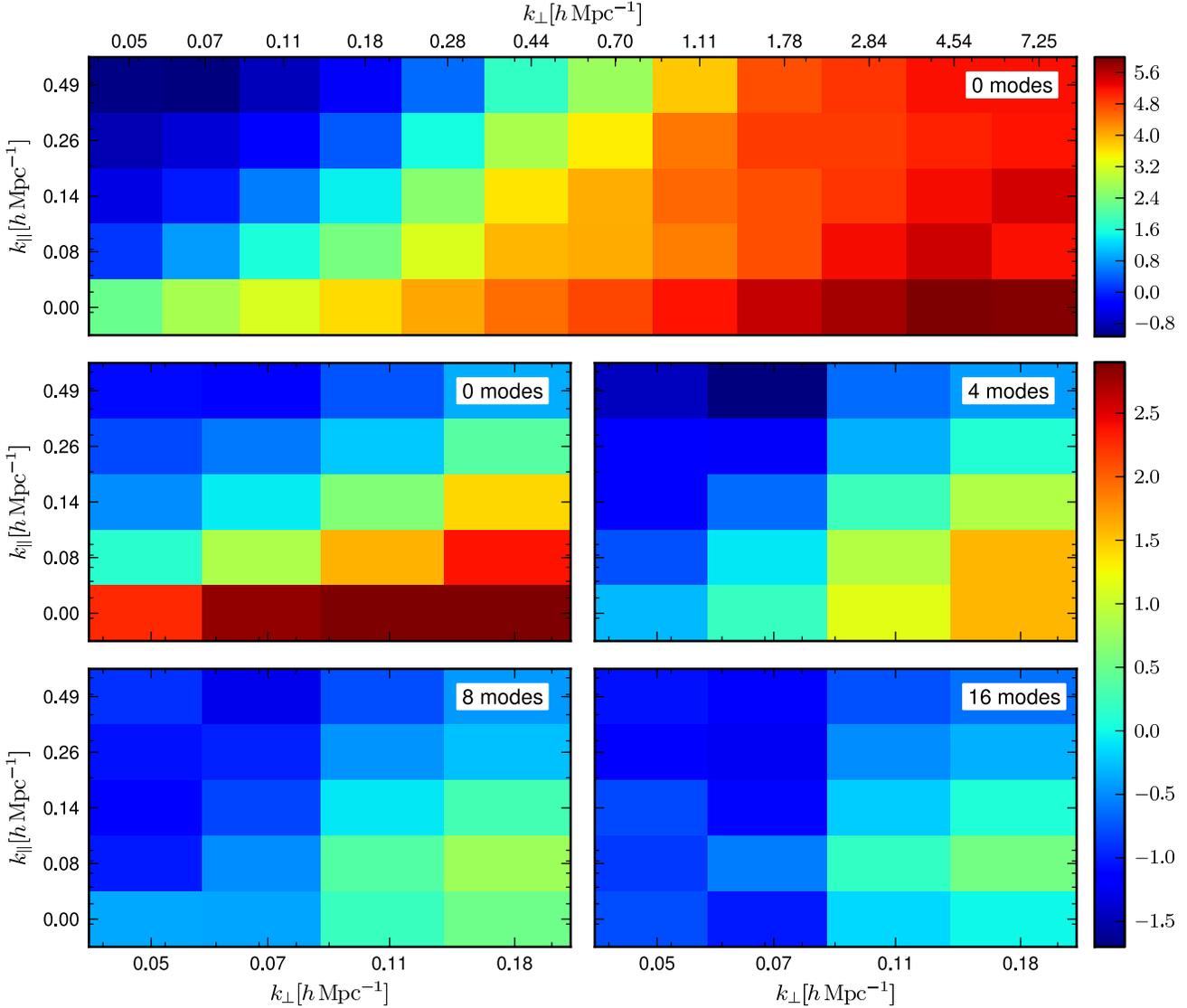}
\caption{Power spectra as a function of both $\kperp$ and $\kpara$
corrected with $T_1$ calculated using the semi-Hermite approach.
The topmost plot shows the entire $\kperp$ range
without any foreground removal.
The four smaller plots show
0, 4, 8, and 16 SVD modes removed on the same colour scale
for only the lowest few $\kperp$ bins.
The colour scales are in units of $\log(\mathrm{K^2})$.
Compared to the case with 0 modes removed, the SVD
tends to reduce the overall power by one to three orders of magnitude.
See also Fig.~\ref{fig:pktotal}, which shows the power
as a function of the total $k$.}
\label{fig:pcolor-t1-single}
\end{figure*}

\begin{figure*}
\includegraphics[width=\textwidth]{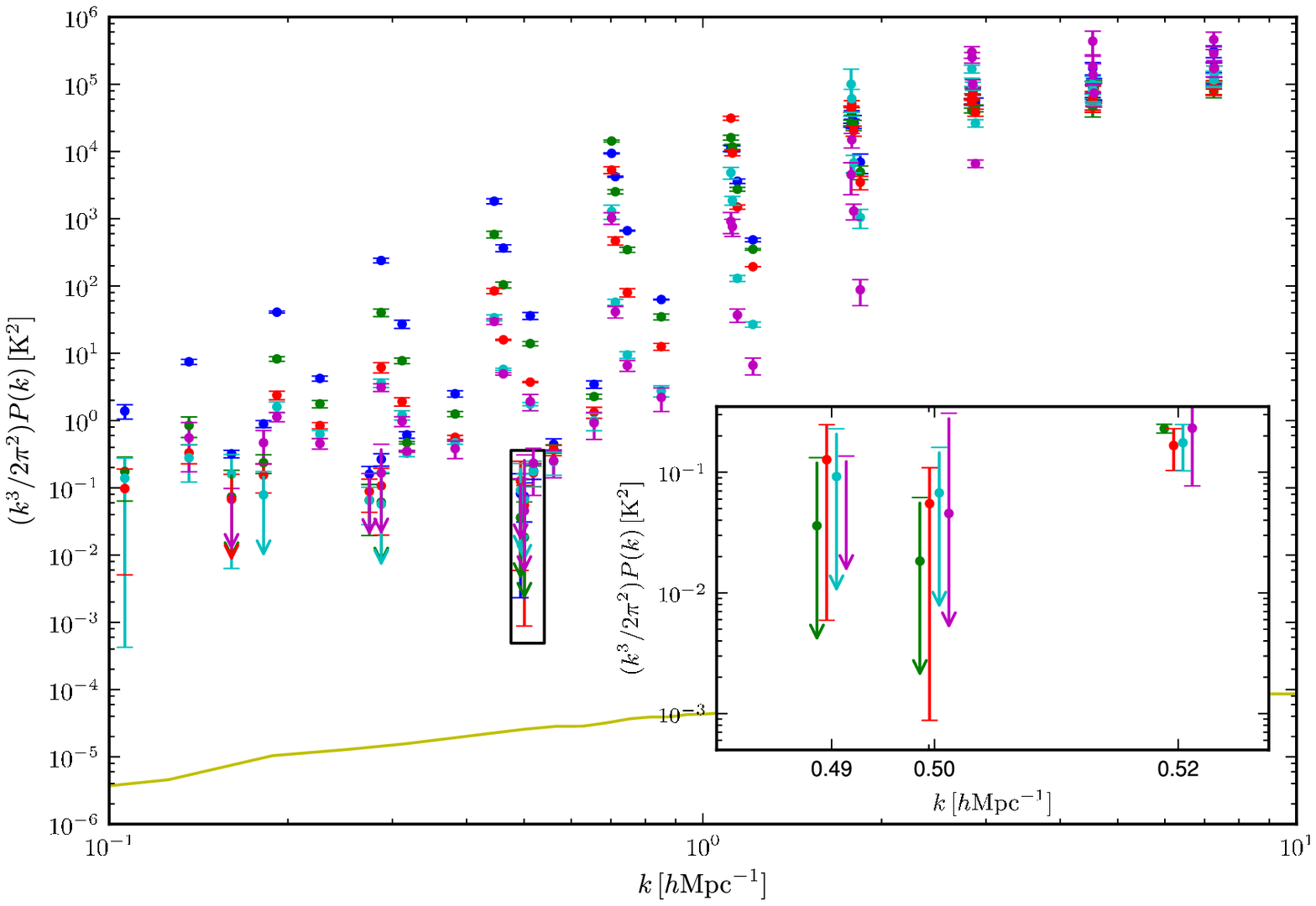}
\caption{Power as a function of the total wavenumber $k = \sqrt{\kperp^2 + \kpara^2}$.
Each point represents a different $(\kperp, \kpara)$ pair; 
there is no binning in $k$.
Colours indicate the number of SVD modes removed;
0 (blue), 4 (green), 8 (red), 16 (cyan), and 32 (purple) are shown.
The boxed region at $k\approx0.5$ is shown inset,
with nearby points each of the three marked $k$ spread out slightly for clarity. 
The best limit at $2\sigma$ 
is (248\,mK)$^2$ at 0.50\,$h$Mpc$^{-1}$
achieved with four SVD modes removed.
The solid line shows the predicted 3D power spectrum from \citet{Iliev08}
assuming a 30\,mK signal.}
\label{fig:pktotal}
\end{figure*}

Fig.~\ref{fig:transfer-example}
shows a typical power spectrum 
for a particular choice of SVD filter, transfer function,
and $\kpara$,
without any correction
and the resulting spectra after each of the above
approaches. 
Differences in each approach illustrate
the difficulty in finding an unbiased estimator
that gives a robust result.

We find that in both the full Hermite and semi-Hermite 
methods there is a $\kpara$ dependence which
is not captured by the SVD-only method,
which is constant with $\kpara$ by definition.
All three methods show deviations from unity of the order of a few per cent 
with zero SVD modes removed
due to the additional effects from RFI masking, noise, and the beam
that are captured by the transfer function.
It is notable, however,
that the `full Hermite' approach
finds $T_1$ deviating from 1 by tens of percent in some regimes,
especially at low $\kperp$.
This is likely indicative of 
a mismatch between the amount of power
being removed by the combination of SVD and Hermite filters
and the amount modelled by the analytic form.
This suggests that in areas of $(\kperp, \kpara)$ space where $T_1>1$,
this method may overestimate the amount of signal present,
in turn underestimating the 21\,cm power
by failing to fully correct for the signal loss.
None the less, the full and semi-Hermite approaches agree
much better with more SVD modes removed.
Since the semi-Hermite approach seems
to capture both the $\kpara$ dependence and is
relatively well behaved with $0<T_1<1$,
we use it as the canonical transfer function.

\subsection{Sampling $(\kperp, \kpara)$ space to get $P(k)$}

Using the Hermite window to select a fixed $\kpara$ allows
us to calculate $P(\ell | \kpara)$
and the associated transfer function at that $\kpara$.
By repeating this for
a series of $\kpara$, we can build up 
the full 3D power spectrum.

Fig.~\ref{fig:pcolor-t1-single} shows the power as 
a function of both $\kperp$ and $\kpara$ using
the semi-Hermite correction, given a series of
different SVD mode subtractions.
The power shows a pattern of lower values towards
low $\kperp$ and high $\kpara$.
Fig.~\ref{fig:pktotal} shows the same measurements
as a function of the 3D wavenumber
$k = \sqrt{\kperp^2 + \kpara^2}$.
Though the SVD is our primary mode of foreground removal,
the Hermite function itself acts as a
foreground filter removing the large-scale structure in frequency space.
This is reflected in the points where zero SVD modes have
been removed.
It is clear that our ability to remove foregrounds
drops off quickly above about $k\approx0.5\,h\,\mathrm{Mpc}^{-1}$.
Our best limit at $2\sigma$
is $(248\,\mathrm{mK})^2$, achieved at
$(\kperp, \kpara) = (0.11,0.49)\,h\,\mathrm{Mpc}^{-1}$,
or a total $k$ of $0.50\,h\,\mathrm{Mpc}^{-1}$,
with four SVD modes removed.
At this point, the semi-Hermite value of the
transfer function was $T_1 = 0.74$,
meaning that an estimated 26 per cent of signal was removed
by the SVD mode subtraction and Hermite window
operating on each day in the cross-correlations.
If instead 16 modes are removed,
the limit changes to
$(319\,\mathrm{mK})^2$
but 55 per cent of the signal is lost.
Any residual foregrounds, though
reduced by a much larger fraction than the signal,
 will also have been boosted by this correction,
making this measurement an upper
limit on the actual 21\,cm signal.

\section{Conclusion}
\label{sec:conclusion}

Using an SVD as a foreground
removal technique and a simulated signal to quantify the
loss of a real 21\,cm signal the SVD may cause,
we have calculated an upper limit to the H{\small I}
power spectrum at $z=8.6$ of (248\,mK)$^2$
at $k = 0.50\,h\,\mathrm{Mpc}^{-1}$.
The $\kperp$ component was found using the median power
in annuli of the $(u,v)$ plane, while a Hermite window
was used to sample the $\kpara$ direction.
This is in contrast to our previous work with
a piecewise-linear filter which operated only in the
frequency direction and carried with it an implicit $\kpara$ window.

This limit is dependent on the method one chooses to 
calculate the transfer function between the real 21\,cm
signal and the observed power. Both the
$\kperp$ and $\kpara$ behaviour of the foreground filter
chosen needs to be taken into account.
While the semi-Hermite method chosen uses a simulated
signal with power in a limited $\kpara$ window,
and may miss interactions between the SVD filter and
the signal over larger $\kpara$ bands, we believe it
to give the most reliable estimate of the transfer function
and a suitably conservative estimate on the final upper limit.

Had we instead used the full Hermite approach described,
this limit would have been $(260\,\mathrm{mK})^2$.
That this second approach gives a similar value 
suggests that this limit is a fairly robust one.
The difference can likely be attributed in part to the
simplifying assumptions necessary when deriving the
analytical Hermite windowing function.
We also consider the current result to be more
robust than that reported previously in \citet{Paciga11}.
While the previous limit was considerably lower,
this can be accounted for by many factors;
the different $k$ scale,
the change in foreground filter,
several minor changes in the analysis pipeline detailed in section~\ref{sec:data}
and most significantly the fact that
this is the first time a transfer function has been used to correct
for signal lost in the foreground filter.
Without such a correction, our best upper limits
with the SVD foreground filter
may have been incorrectly reported as low as (50\,mK)$^2$.

This limit still compares favourably to others established
in the literature which are of the order of several Kelvin
\citep[e.g.,][]{Bebbington86, Ali08, Parsons10}.
Recently, after submission of our paper, PAPER \citep{Parsons13} claimed an upper limit of
(52\,mK)$^2$ at $k=0.11\,h\,\mathrm{Mpc}^{-1}$ and $z=7.7$.
However, it is not documented whether signal loss from their primary foreground filtering
step (their section 3.4) has been accounted for and so it is not clear how to compare their result 
to ours.
LOFAR has begun publishing initial results from reionization
observations, but have so far focused on much longer scales
($\ell \approx 7500$) \citep{Yatawatta13}.

In \citet{Paciga11} we considered a model with a cold intergalactic medium (IGM),
a neutral fraction of 0.5
and fully ionized bubbles with uniform radii.
In such a model this current limit would constrain
the brightness temperature of the neutral IGM
to be at least 540\,mK in absorption against the CMB.
However,
a value of the H{\small I} power spectrum of (248\,mK)$^2$
is almost an order of magnitude higher than what is generally
considered physically plausible
in most reionization models.
In particular, this result does not constrain reionization models
with a warm IGM where the spin temperature is much greater 
than the CMB temperature.

The SVD procedure could be refined further by a
baseline-by-baseline accounting of the optimal
number of modes to subtract
or by limiting the field of view on the sky to the innermost
area of the beam where point source residuals are minimal,
although it is not obvious what effect this would have on
the signal at small angular scales.
Making a measurement at larger $\ell$ would
require a more careful treatment of point sources
but is also limited by the fact that the SVD
is less effective for longer baselines.
Regardless of the foreground removal technique used,
it is likely that accurately correcting for 
the any resulting loss of the 21\,cm signal,
and disentangling the 21\,cm signal from
any residual foregrounds, will remain a significant
challenge in measuring the true EoR power spectrum.

\section{Acknowledgements}
\label{sec:Acknowledgments}

Special thanks to Eric Switzer and Kiyoshi Masui for many useful discussions
during the preparation of this work.
We thank the staff of the GMRT that made these observations
possible. GMRT is run by the National Centre for Radio Astrophysics of
the Tata Institute of Fundamental Research.
The computations were performed on CITA's Sunnyvale clusters
which are funded by the Canada Foundation for Innovation,
the Ontario Innovation Trust
and the Ontario Research Fund.
The work of GP, UP and KS is supported by the
Natural Science and Engineering Research Council of Canada.
KB, JP and TV are supported by NSF grant AST-1009615.

\bibliography{svdpaper}

\begin{thebibliography}{}

\bibitem[\protect\citeauthoryear{{Ali}, {Bharadwaj} \& {Chengalur}}{{Ali}
  et~al.}{2008}]{Ali08}
{Ali} S.~S.,  {Bharadwaj} S.,    {Chengalur} J.~N.,  2008, MNRAS, 385, 2166

\bibitem[\protect\citeauthoryear{{Ananthakrishnan}}{{Ananthakrishnan}}{1995}]{%
GMRT}
{Ananthakrishnan} S.,  1995, Journal of Astrophysics and Astronomy Supplement,
  16, 427

\bibitem[\protect\citeauthoryear{{Beardsley} et~al.}{{Beardsley} et~al.}{2013}]{Beardsley12}
{Beardsley} A.~P. {et al.}, 2013,
  MNRAS, 429, L5

\bibitem[\protect\citeauthoryear{{Bebbington}}{{Bebbington}}{1986}]{Bebbington%
86}
{Bebbington} D.~H.~O.,  1986, MNRAS, 218, 577

\bibitem[\protect\citeauthoryear{{Becker} et~al.}{{Becker}
  et~al.}{2001}]{Becker01}
{Becker} R.~H. {et al.} 2001, AJ, 122, 2850

\bibitem[\protect\citeauthoryear{{Bernardi}, {Mitchell}, {Ord}, {Greenhill},
  {Pindor}, {Wayth} \& {Wyithe}}{{Bernardi} et~al.}{2011}]{Bernardi11}
{Bernardi} G.,  {Mitchell} D.~A.,  {Ord} S.~M.,  {Greenhill} L.~J.,  {Pindor}
  B.,  {Wayth} R.~B.,    {Wyithe} J.~S.~B.,  2011, MNRAS, 413, 411

\bibitem[\protect\citeauthoryear{{Bharadwaj} \& {Ali}}{{Bharadwaj} \&
  {Ali}}{2005}]{Bharadwaj05}
{Bharadwaj} S.,  {Ali} S.~S.,  2005, MNRAS, 356, 1519

\bibitem[\protect\citeauthoryear{{Bowman} \& {Rogers}}{{Bowman} \&
  {Rogers}}{2010}]{Bowman10}
{Bowman} J.~D.,  {Rogers} A.~E.~E.,  2010, Nature, 468, 796

\bibitem[\protect\citeauthoryear{{Bowman}, {Rogers} \& {Hewitt}}{{Bowman}
  et~al.}{2008}]{Bowman08}
{Bowman} J.~D.,  {Rogers} A.~E.~E.,    {Hewitt} J.~N.,  2008, ApJ, 676, 1

\bibitem[\protect\citeauthoryear{{Brentjens}, {Koopmans}, {de Bruyn} \&
  {Zaroubi}}{{Brentjens} et~al.}{2011}]{Brentjens11}
{Brentjens} M.,  {Koopmans} L.~V.~E.,  {de Bruyn} A.~G.,    {Zaroubi} S.,
  2011, in American Astronomical Society Meeting Abstracts \#217 Vol.~43 of
  Bulletin of the American Astronomical Society, {The Low-Frequency Array
  (LOFAR) and EoR Key-Science Project}.
p. \#107.04

\bibitem[\protect\citeauthoryear{{Carilli}, {Furlanetto}, {Briggs}, {Jarvis},
  {Rawlings} \& {Falcke}}{{Carilli} et~al.}{2004}]{Carilli04}
{Carilli} C.~L.,  {Furlanetto} S.,  {Briggs} F.,  {Jarvis} M.,  {Rawlings} S.,
    {Falcke} H.,  2004, New~Astron.~Rev., 48, 1029

\bibitem[\protect\citeauthoryear{{Chang}, {Pen}, {Bandura} \&
  {Peterson}}{{Chang} et~al.}{2010}]{Chang10}
{Chang} T.-C.,  {Pen} U.-L.,  {Bandura} K.,    {Peterson} J.~B.,  2010, Nature,
  466, 463

\bibitem[\protect\citeauthoryear{{Chapman}, {Abdalla}, {Harker}, {Jeli{\'c}},
  {Labropoulos}, {Zaroubi}, {Brentjens}, {de Bruyn} \& {Koopmans}}{{Chapman}
  et~al.}{2012}]{Chapman12}
{Chapman} E. el~al.,  2012, MNRAS, 423, 2518

\bibitem[\protect\citeauthoryear{{Cooray}, {Li} \& {Melchiorri}}{{Cooray}
  et~al.}{2008}]{Cooray08}
{Cooray} A.,  {Li} C.,    {Melchiorri} A.,  2008, Phys.~Rev.~D, 77, 103506

\bibitem[\protect\citeauthoryear{{Datta}, {Bowman} \& {Carilli}}{{Datta}
  et~al.}{2010}]{Datta10}
{Datta} A.,  {Bowman} J.~D.,    {Carilli} C.~L.,  2010, ApJ, 724, 526

\bibitem[\protect\citeauthoryear{{Datta}, {Friedrich}, {Mellema}, {Iliev} \&
  {Shapiro}}{{Datta} et~al.}{2012}]{Datta12}
{Datta} K.~K.,  {Friedrich} M.~M.,  {Mellema} G.,  {Iliev} I.~T.,    {Shapiro}
  P.~R.,  2012, MNRAS, 424, 762

\bibitem[\protect\citeauthoryear{{de Oliveira-Costa}, {Tegmark}, {Gaensler},
  {Jonas}, {Landecker} \& {Reich}}{{de Oliveira-Costa} et~al.}{2008}]{dOC08}
{de Oliveira-Costa} A.,  {Tegmark} M.,  {Gaensler} B.~M.,  {Jonas} J.,
  {Landecker} T.~L.,    {Reich} P.,  2008, MNRAS, 388, 247

\bibitem[\protect\citeauthoryear{{Dillon}, {Liu} \& {Tegmark}}{{Dillon}
  et~al.}{2013}]{Dillon12}
{Dillon} J.~S.,  {Liu} A.,    {Tegmark} M.,  2013, Phys.~Rev.D, 87, 43005

\bibitem[\protect\citeauthoryear{{Djorgovski}, {Castro}, {Stern} \&
  {Mahabal}}{{Djorgovski} et~al.}{2001}]{Djorgovski01}
{Djorgovski} S.~G.,  {Castro} S.,  {Stern} D.,    {Mahabal} A.~A.,  2001, ApJl,
  560, L5

\bibitem[\protect\citeauthoryear{{Friedrich}, {Mellema}, {Alvarez}, {Shapiro}
  \& {Iliev}}{{Friedrich} et~al.}{2011}]{Friedrich11}
{Friedrich} M.~M.,  {Mellema} G.,  {Alvarez} M.~A.,  {Shapiro} P.~R.,
  {Iliev} I.~T.,  2011, MNRAS, 413, 1353

\bibitem[\protect\citeauthoryear{{Furlanetto} et~al.}{{Furlanetto} et~al.}{2009}]{Furlanetto09}
{Furlanetto} S.~R. {et al.} 2009, Astro2010: The
  Astronomy and Astrophysics Decadal Survey, Science White Papers, 82

\bibitem[\protect\citeauthoryear{{Furlanetto}, {Oh} \& {Briggs}}{{Furlanetto}
  et~al.}{2006}]{Furlanetto06}
{Furlanetto} S.~R.,  {Oh} S.~P.,    {Briggs} F.~H.,  2006, Phys.~Rep., 433, 181

\bibitem[\protect\citeauthoryear{{Furlanetto}, {Oh} \&
  {Pierpaoli}}{{Furlanetto} et~al.}{2006}]{Furlanetto06BHA}
{Furlanetto} S.~R.,  {Oh} S.~P.,    {Pierpaoli} E.,  2006, Phys.~Rev.~D, 74,
  103502

\bibitem[\protect\citeauthoryear{{Furlanetto}, {Zaldarriaga} \&
  {Hernquist}}{{Furlanetto} et~al.}{2004}]{Furlanetto04}
{Furlanetto} S.~R.,  {Zaldarriaga} M.,    {Hernquist} L.,  2004, ApJ, 613, 1

\bibitem[\protect\citeauthoryear{{Ghosh}, {Prasad}, {Bharadwaj}, {Ali} \&
  {Chengalur}}{{Ghosh} et~al.}{2012}]{Ghosh12}
{Ghosh} A.,  {Prasad} J.,  {Bharadwaj} S.,  {Ali} S.~S.,    {Chengalur} J.~N.,
  2012, MNRAS, 426, 3295

\bibitem[\protect\citeauthoryear{{Griffen}, {Drinkwater}, {Iliev}, {Thomas} \&
  {Mellema}}{{Griffen} et~al.}{2013}]{Griffen12}
{Griffen} B.~F.,  {Drinkwater} M.~J.,  {Iliev} I.~T.,  {Thomas} P.~A.,
  {Mellema} G.,  2013, MNRAS, 431, 3087

\bibitem[\protect\citeauthoryear{{Gunn} \& {Peterson}}{{Gunn} \&
  {Peterson}}{1965}]{GunnPeterson65}
{Gunn} J.~E.,  {Peterson} B.~A.,  1965, ApJ, 142, 1633

\bibitem[\protect\citeauthoryear{{Haiman}}{{Haiman}}{2011}]{Haiman11}
{Haiman} Z.,  2011, Nature, 472, 47

\bibitem[\protect\citeauthoryear{{Harker}, {Zaroubi}, {Bernardi}, {Brentjens},
  {de Bruyn}, {Ciardi}, {Jeli{\'c}}, {Koopmans}, {Labropoulos}, {Mellema},
  {Offringa}, {Pandey}, {Pawlik}, {Schaye}, {Thomas} \& {Yatawatta}}{{Harker}
  et~al.}{2010}]{Harker10}
{Harker} G. et~al.,  2010, MNRAS, 405, 2492

\bibitem[\protect\citeauthoryear{{Iliev}, {Mellema}, {Pen}, {Bond} \&
  {Shapiro}}{{Iliev} et~al.}{2008}]{Iliev08}
{Iliev} I.~T.,  {Mellema} G.,  {Pen} U.-L.,  {Bond} J.~R.,    {Shapiro} P.~R.,
  2008, MNRAS, 384, 863

\bibitem[\protect\citeauthoryear{{Iliev}, {Mellema}, {Shapiro}, {Pen}, {Mao},
  {Koda} \& {Ahn}}{{Iliev} et~al.}{2012}]{Iliev12}
{Iliev} I.~T.,  {Mellema} G.,  {Shapiro} P.~R.,  {Pen} U.-L.,  {Mao} Y.,
  {Koda} J.,    {Ahn} K.,  2012, MNRAS, 423, 2222

\bibitem[\protect\citeauthoryear{{Jeli{\'c}}, {Zaroubi}, {Labropoulos},
  {Thomas}, {Bernardi}, {Brentjens}, {de Bruyn}, {Ciardi}, {Harker},
  {Koopmans}, {Pandey}, {Schaye} \& {Yatawatta}}{{Jeli{\'c}}
  et~al.}{2008}]{Jelic08}
{Jeli{\'c}} V. et~al.,  2008,
  MNRAS, 389, 1319

\bibitem[\protect\citeauthoryear{{Komatsu}, {Smith}, {Dunkley} \& {et
  al.}}{{Komatsu} et~al.}{2011}]{Komatsu11}
{Komatsu} E.,  {Smith} K.~M.,  {Dunkley} J.,    {et al.} 2011, ApJS, 192, 18

\bibitem[\protect\citeauthoryear{{Kovetz} \& {Kamionkowski}}{{Kovetz} \&
  {Kamionkowski}}{2013}]{Kovetz12}
{Kovetz} E.~D.,  {Kamionkowski} M.,  2013, Phys.~Rev.~D, 87, 63516

\bibitem[\protect\citeauthoryear{{Liu} \& {Tegmark}}{{Liu} \&
  {Tegmark}}{2011}]{Liu11}
{Liu} A.,  {Tegmark} M.,  2011, Phys.~Rev.~D, 83, 103006

\bibitem[\protect\citeauthoryear{{Liu} \& {Tegmark}}{{Liu} \&
  {Tegmark}}{2012}]{Liu12}
{Liu} A.,  {Tegmark} M.,  2012, MNRAS, 419, 3491

\bibitem[\protect\citeauthoryear{{Lonsdale}, {Cappallo}, {Morales} \& {et
  al.}}{{Lonsdale} et~al.}{2009}]{Lonsdale09}
{Lonsdale} C.~J.,  {Cappallo} R.~J.,  {Morales} M.~F.,    {et al.} 2009, IEEE
  Proceedings, 97, 1497

\bibitem[\protect\citeauthoryear{{Majumdar}, {Bharadwaj} \&
  {Choudhury}}{{Majumdar} et~al.}{2012}]{Majumdar12}
{Majumdar} S.,  {Bharadwaj} S.,    {Choudhury} T.~R.,  2012, MNRAS, 426, 3178

\bibitem[\protect\citeauthoryear{{Mao}, {Tegmark}, {McQuinn}, {Zaldarriaga} \&
  {Zahn}}{{Mao} et~al.}{2008}]{Mao08}
{Mao} Y.,  {Tegmark} M.,  {McQuinn} M.,  {Zaldarriaga} M.,    {Zahn} O.,  2008,
  Phys.~Rev.~D, 78, 023529

\bibitem[\protect\citeauthoryear{{Masui}, {Switzer}, {Banavar}, {Bandura},
  {Blake}, {Calin}, {Chang}, {Chen}, {Li}, {Liao}, {Natarajan}, {Pen},
  {Peterson}, {Shaw} \& {Voytek}}{{Masui} et~al.}{2012}]{Masui12}
{Masui} K.~W. et~al.,  2013, ApJL, 763, L20

\bibitem[\protect\citeauthoryear{{McGreer}, {Mesinger} \& {Fan}}{{McGreer}
  et~al.}{2011}]{McGreer11}
{McGreer} I.~D.,  {Mesinger} A.,    {Fan} X.,  2011, MNRAS, 415, 3237

\bibitem[\protect\citeauthoryear{{McQuinn}, {Lidz}, {Zahn}, {Dutta},
  {Hernquist} \& {Zaldarriaga}}{{McQuinn} et~al.}{2007}]{McQuinn07}
{McQuinn} M.,  {Lidz} A.,  {Zahn} O.,  {Dutta} S.,  {Hernquist} L.,
  {Zaldarriaga} M.,  2007, MNRAS, 377, 1043

\bibitem[\protect\citeauthoryear{{McQuinn}, {Zahn}, {Zaldarriaga}, {Hernquist}
  \& {Furlanetto}}{{McQuinn} et~al.}{2006}]{McQuinn06}
{McQuinn} M.,  {Zahn} O.,  {Zaldarriaga} M.,  {Hernquist} L.,    {Furlanetto}
  S.~R.,  2006, ApJ, 653, 815

\bibitem[\protect\citeauthoryear{{Nityananda}}{{Nityananda}}{2010}]{Nityananda%
10}
{Nityananda} R.,  2010, {NCRA Technical Report,
  \url{http://ncralib1.ncra.tifr.res.in:8080/jspui/handle/2301/484}}

\bibitem[\protect\citeauthoryear{{Oh} \& {Mack}}{{Oh} \& {Mack}}{2003}]{Oh03}
{Oh} S.~P.,  {Mack} K.~J.,  2003, MNRAS, 346, 871

\bibitem[\protect\citeauthoryear{{Paciga}, {Chang}, {Gupta}, {Nityananda},
  {Odegova}, {Pen}, {Peterson}, {Roy} \& {Sigurdson}}{{Paciga}
  et~al.}{2011}]{Paciga11}
{Paciga} G. et~al.,  2011, MNRAS,
  413, 1174

\bibitem[\protect\citeauthoryear{{Pandolfi}, {Ferrara}, {Choudhury},
  {Melchiorri} \& {Mitra}}{{Pandolfi} et~al.}{2011}]{Pandolfi11}
{Pandolfi} S.,  {Ferrara} A.,  {Choudhury} T.~R.,  {Melchiorri} A.,    {Mitra}
  S.,  2011, Phys.~Rev.~D, 84, 123522

\bibitem[\protect\citeauthoryear{{Parsons}, {Pober}, {McQuinn}, {Jacobs} \&
  {Aguirre}}{{Parsons} et~al.}{2012}]{Parsons11}
{Parsons} A.,  {Pober} J.,  {McQuinn} M.,  {Jacobs} D.,    {Aguirre} J.,  2012,
  ApJ, 753, 81

\bibitem[\protect\citeauthoryear{{Parsons}, {Backer}, {Foster}, {Wright},
  {Bradley}, {Gugliucci}, {Parashare}, {Benoit}, {Aguirre}, {Jacobs},
  {Carilli}, {Herne}, {Lynch}, {Manley} \& {Werthimer}}{{Parsons}
  et~al.}{2010}]{Parsons10}
{Parsons} A.~R. et~al.,  2010, AJ, 139, 1468

\bibitem[\protect\citeauthoryear{{Parsons}, {Pober}, {Aguirre}, {Carilli},
  {Jacobs} \& {Moore}}{{Parsons} et~al.}{2012}]{Parsons12}
{Parsons} A.~R.,  {Pober} J.~C.,  {Aguirre} J.~E.,  {Carilli} C.~L.,  {Jacobs}
  D.~C.,    {Moore} D.~F.,  2012, ApJ, 756, 165

\bibitem[\protect\citeauthoryear{{Parsons} et~al.}{{Parsons} et~al.}{2013}]{Parsons13}
{Parsons} A.~R. et~al., 2013, preprint (astro-ph/astro-ph/1208.0331v2)

\bibitem[\protect\citeauthoryear{{Petrovic} \& {Oh}}{{Petrovic} \&
  {Oh}}{2011}]{Petrovic11}
{Petrovic} N.,  {Oh} S.~P.,  2011, MNRAS, 413, 2103

\bibitem[\protect\citeauthoryear{{Rawlings} \& {Schilizzi}}{{Rawlings} \&
  {Schilizzi}}{2011}]{Rawlings11}
{Rawlings} S.,  {Schilizzi} R.,  2011, preprint (astro-ph/1105.5953)

\bibitem[\protect\citeauthoryear{{Schroeder}, {Mesinger} \&
  {Haiman}}{{Schroeder} et~al.}{2012}]{Schroeder12}
{Schroeder} J.,  {Mesinger} A.,    {Haiman} Z.,  2012, MNRAS, p.~203

\bibitem[\protect\citeauthoryear{{Su}, {Yadav}, {McQuinn}, {Yoo} \&
  {Zaldarriaga}}{{Su} et~al.}{2011}]{Su11}
{Su} M.,  {Yadav} A.~P.~S.,  {McQuinn} M.,  {Yoo} J.,    {Zaldarriaga} M.,
  2011, preprint (astro-ph/1106.4313)

\bibitem[\protect\citeauthoryear{{Swarup}, {Ananthakrishnan}, {Kapahi}, {Rao},
  {Subrahmanya} \& {Kulkarni}}{{Swarup} et~al.}{1991}]{Swarup91}
{Swarup} G.,  {Ananthakrishnan} S.,  {Kapahi} V.~K.,  {Rao} A.~P.,
  {Subrahmanya} C.~R.,    {Kulkarni} V.~K.,  1991, Current Science, Vol.~60,
  NO.2/JAN25, P.~95, 1991, 60, 95

\bibitem[\protect\citeauthoryear{{Trott}, {Wayth} \& {Tingay}}{{Trott}
  et~al.}{2012}]{Trott12}
{Trott} C.~M.,  {Wayth} R.~B.,    {Tingay} S.~J.,  2012, ApJ, 757, 101

\bibitem[\protect\citeauthoryear{{Yatawatta} et~al.}{{Yatawatta}
  et~al.}{2013}]{Yatawatta13}
{Yatawatta} S. et~al., 2013, A\&A, 550, A136

\bibitem[\protect\citeauthoryear{{Zahn}, {Lidz}, {McQuinn} \& {Dutta}}{{Zahn}
  et~al.}{2007}]{Zahn07}
{Zahn} O.,  {Lidz} A.,  {McQuinn} M.,    {Dutta} S.,  2007, ApJ, 654, 12

\bibitem[\protect\citeauthoryear{{Zahn}, {Reichardt}, {Shaw} \& {et
  al.}}{{Zahn} et~al.}{2012}]{Zahn12}
{Zahn} O.,  {Reichardt} C.~L.,  {Shaw} L.,    {et al.} 2012, ApJ, 756, 65

\bibitem[\protect\citeauthoryear{{Zaroubi}, {de Bruyn}, {Harker} \& {et
  al.}}{{Zaroubi} et~al.}{2012}]{Zaroubi12}
{Zaroubi} S.,  {de Bruyn} A.~G.,  {Harker} G.,    {et al.} 2012, MNRAS, 425,
  2964

\end{thebibliography}

\end{document}